\documentclass[aps,prb,twocolumn,superscriptaddress,showpacs,floatfix]{revtex4}
\usepackage{amsmath}
\usepackage{graphicx}
\usepackage[usenames]{color}

\def\bra#1{\mathinner{\langle{#1}|}}
\def\ket#1{\mathinner{|{#1}\rangle}}
\newcommand{\braket}[2]{\langle #1|#2\rangle}
\newcommand{\ketbra}[2]{|#1 \rangle \langle #2|}
\def\up{\uparrow}
\def\down{\downarrow}

\begin{document}
\title{Entanglement evolution of a spin chain bath in driving the decoherence of a coupled quantum spin}

\author{Zhi-Hui Wang}
\affiliation{Institute of Theoretical Physics, Chinese Academy of
Sciences, Beijing 100190, China}

\author{Bing-Shen Wang}
\affiliation{State Key Laboratory of Semiconductor Superlattice
and Microstructure and Institute of Semiconductor, Chinese Academy
of Sciences, Beijing 100083, China}

\author{Zhao-Bin Su}
\affiliation{Institute of Theoretical Physics, Chinese Academy of
Sciences, Beijing 100190, China}

\date{\today}
\pacs{03.65.Yz, 03.67.Mn, 75.10.Jm, 03.65.Ta }

\begin{abstract}
For an electron spin in coupling with an interacting spin chain via
hyperfine-type interaction, we investigate the dynamical evolutions
of the pairwise entanglement of the spin chain and a correlation
function joined the electron spin with a pair of chain spins in
correspondence to the electron spin coherence evolution. Both
quantities manifest a periodic and a decaying evolution. The
entanglement of the spin bath is significant in distinguishing the
zero-coherence status exhibited in periodic and decoherence
evolutions of the electron spin. The periodical concurrence
evolution of the spin bath characterizes the whole system in a
coherence-preserving phase, particularly for the case that the
associated periodic coherence evolution is predominated by
zero-value in the infinite chain-length limit, which was often
regarded as the realization of decoherence.
\end{abstract}
\maketitle

Quantum coherence in terms of state superposition is one of the most
important features of quantum mechanics. Usually the unavoidable
environment would render the quantum object decohered. In most cases
the environment can be efficiently modeled by an infinite number of
non-interacting oscillators, i.e., the boson bath as in the
Caldeira-Legget model.\cite{C-L} The decohering environment can also
be a spin bath. An interesting example is the so-called Coleman-Hepp
model,\cite{Hepp} proposed in the study of quantum measurement. In
this model, the apparatus is an one-dimensional chain with infinite
number of non-interacting spins, which is used to measure and to
collapse the superposed spin states of a relativistic electron.

The decoherence of an electron spin $S$ induced by an environment
$E$ can be viewed as a quantum measurement process. The electron
spin in a state $\ket{\phi^S}=c_1\ket{+}+c_2\ket{-}$ is to be
measured by $E$ as the apparatus. The quantum dynamics of the
universe, i.e., the closed system formed by $S$ and $E$, is expected
to result in the von Neumann's wave packet collapse
postulate\cite{von-Neumann, Sun_1993}
\begin{eqnarray*}
\rho^S(t) = Tr_E[\rho(t)]\rightarrow |c_1|^2\ketbra{+}{+}
+|c_2|^2\ketbra{-}{-}\;,
\end{eqnarray*}
where $\rho^S(t)$ is the reduced density matrix of $S$,
and $\rho(t)$ the evolving density matrix of the universe.
The decoherence of $S$ is actually the realization of
wave packet collapse in sense of von Neumann's postulate.

Since the universe is composed of two mutually interacting parts,
$S$ and $E$, the bath-traced electron-spin coherence should be
complemented by a proper description of the environment status. The
relation between the entanglement of the bath and the coherence of
the system has stimulated extensive interest.\cite{Sun_2001,Milburn}
However, the dynamical evolution of the coherence status of the
environment in correspondence to the coherence evolution (CE) of $S$
has not been systematically explored. In this paper, we investigate
the entanglement evolution of a spin chain bath in driving the
decohrence of a coupled electron spin. In fact, zero-coherence
status can be realized in two kinds of qualitatively different
electron spin CE's, which exhibit different responses to the spin
echo effect.\cite{Wang} It is then desirable to explore the
underlying difference between these two coherence states and its
implications by studying the status of the environment. 

In this paper, we introduce the pairwise concurrence which measures
the entanglement of the spin chain, and a joint-correlation function
which correlates the system spin with a pair of bath spins. We find
that, the evolutions of both quantities exhibit critically different
behaviors in correspondence to the different CE of $S$. The apparent
zero-coherence status appeared in the two kinds of CE's prevail in
the joint-correlation evolution but can be discriminated by the
concurrence of the spin chain.
In particular, corresponding to the periodic coherence evolution
predominated by zero values, the bath chain maintains in a stable
entangled state with periodic concurrence evolution while the
zero-coherence status appeared in the decoherence evolution
corresponds to the bath chain with disentangled spin pairs.
Moreover, the periodicity of bath concurrence persists even when the
coherence exhibits a non-decaying irregularly oscillating evolution.
The periodic concurrence evolution of the spin bath with non-zero
amplitude characterizes the whole system in a coherence-preserving
phase.

{\sl Model description.---}
The Hamiltonian for the universe
\begin{eqnarray}
\label{eq:Hami}
H=H_S+H_E+H_I
\end{eqnarray}
is composed of the system part $H_S = \epsilon_+ \ketbra{+}{+} +
\epsilon_- \ketbra{-}{-}$, the environment part $H_E =
B/{2}\sum_{j=1}^{N-1} \big(I^+_j I^-_{j+1} + I^-_jI^+_{j+1}\big)$
and the interaction part $H_I = \sum_{j=1}^{N} A_j S^z I^z_j$, which
is the longitudinal hyperfine (HF)-type interaction\cite{Loss}
between $S$ and $E$. Here $j$ is the site index of the spin chain,
$I_j^\pm=I_j^x\pm iI_j^y$, and ${\bf I}_j$ the corresponding spin
operator with $I=\frac{1}{2}$. $\epsilon_\pm$ are the Zeeman
energies of the electron spin under applied magnetic field, $B$ the
coupling constant of the nearest-neighbored chain spins, and $A_j$
the HF-type interaction strength between $S$ and $E$, which varies
from site to site on the spin chain.

We notice that, the z-component of the total bath spin $I^z=\sum_j
I^z_j$ is a constant of motion of the whole system, $[I^z,H]=0$. It
is then physically reasonable to confine the bath states to the
$I^z=0$ section of the full Hilbert space. As a result, the
contribution of the longitudinal HF-type interaction depends only on
the differences of the local interaction strength, i.e., it keeps
unchanged with the substitution $A_j\to A_j+ $common const. for all
sites.\cite{Wang} Moreover, the z-component of the electron spin
$S^z$ is also a good quantum number and the total Hamiltonian is
diagonal in the $S^z$ representation as
\begin{eqnarray}
H=\ketbra{+}{+} H_+ + \ketbra{-}{-} H_- \end{eqnarray}
with
electron-spin-conditioned bath Hamiltonian
\begin{eqnarray}
\label{eq:Hami_mapped} H_\pm = \epsilon_\pm + H_E \pm
\sum_j\frac{A_j}{2} I^z_j \;.
\end{eqnarray}

We consider two kinds of HF-type interactions: (i) the linear form $
A_j\mbox{=}(N-j)\Delta A\mbox{+}A_N $ with $\Delta A$ being a
constant, and the resulted CE depending only on the ratio of the
slope of the HF-type interaction and the intra-bath interaction
strength, $\xi=\frac{\Delta A}{B}$. \cite{Wang} (ii) the cosine form
$A_j=\mathcal A \cos^2\frac{(j-1)\pi}{2N}$ with magnitude $\cal A$
constant, which resembles the HF coupling in a quantum dot with
$A_j$ proportional to the norm of the electron wave
function.\cite{Sham_DasSarma} The former is a minimal model to
clarify the underlying mechanism for decoherence while the latter is
more realistic. 

At time $t=0$, $S$ and $E$ are in pure states
and are disentangled from each other as
$
\rho_0 = \rho^S_0\otimes\rho^E_0
= \big(\ketbra{\phi^S}{\phi^S}\big)\otimes \big{(\ketbra{\Phi^E}{\Phi^E}\big)}
$,
where $\ket{\phi^S}=c_1\ket{+}+c_2\ket{-}$ and $\ket{\Phi^E}$
are the initial states of $S$ and $E$, respectively.
Without loss of generality, we take the initial bath
state randomly selected and fully polarized as
$
\ket{\Phi^E}=\ket{\up\up\down\down\cdots}_{\{j_\alpha\}}\;
$,
where $\{j_\alpha\}$ denotes the initial bath spin configuration
with $M$ spin up states on sites $j_1,j_2,\cdots,j_\alpha,\cdots,j_M$
and spin down states on other sites.
At time $t$, the density matrix of the universe is
$\rho(t)=U(t)\rho_0U^\dagger(t)$
with $U(t)$ the evolution operator of the universe.
In the reduced density matrix of the system
\begin{eqnarray}
\label{eq:rdm}
\rho^S(t)
&=&
|c_1|^2\ketbra{+}{+}
+|c_2|^2\ketbra{-}{-} \nonumber\\
& &
+ c_1c_2^* \ketbra{+}{-}\rho^S_{+,-}(t)
+ h.c. \;,
\end{eqnarray}
the off-diagonal element
\begin{eqnarray}
\label{eq:CE}
\rho^S_{+,-}(t)=Tr[\rho(t)S^-]=\braket{\psi_-(t)}{\psi_+(t)}
\end{eqnarray}
measures the coherence of $S$, where $\ket{\psi_\pm(t)}=e^{-iH_\pm
t}\ket{\Phi^E}$ is the wave function of the spin chain conditioned
on the electron spin states.

Under the Jordan-Wigner transformation,\cite{J-W} $ I^+_j =
c^\dagger_j e^{i\pi\sum_{l=1}^{j-1}c^\dagger_lc_l} $, $ I^-_j$ its
hermitian conjugate and $ I^z_j = c^\dagger_jc_j-\frac{1}{2} $, with
$c^\dagger_j$, $c_j$ the creation and annihilation spinless
fermionic operators on site $j$, the initial bath state reads
$\ket{\Phi^E}=\Pi_{j_\alpha} c^\dagger_{j_\alpha} \ket 0 \;$, the
Hamiltonian $H_\pm$ and the coherence $\rho^S_{+,-}(t)$ take the
forms
\begin{eqnarray}
H_\pm &=& \tilde \epsilon_\pm + \sum^N_{j,j'=1} c^\dagger_j h^\pm_{j,j'}c_{j'}\nonumber\\
\rho^S_{+,-}(t) &=& e^{i(\tilde\epsilon_--\tilde\epsilon_+)t}
\det\big[\chi(t)_{j_\beta,j_{\beta'}}\big]\;,
\end{eqnarray}
where
$h^\pm_{j,j'}=\pm\frac{A_j}{2}\delta_{j,j'}
+\frac{B}{2}(\delta_{j,j'+1}+\delta_{j,j'-1})$,
$\tilde \epsilon_\pm = \epsilon_\pm \mp \frac{1}{4}\sum_j A_j$,
and
$\chi(t)_{j_\beta,j_{\beta'}}=[e^{ih^-t}e^{-ih^+t}]_{j_\beta,j_{\beta'}}$ with
$j_\beta,j_{\beta'} \in \{j_\alpha\}$.

{\sl Joint-correlation function.---} We consider first the joint
correlation function
$G_{j_1,j_2}(t)=Tr[\rho(t)S^+I^+_{j_1}I^-_{j_2}]
=\bra{\psi_+(t)}I^+_{j_1}I^-_{j_2}\ket{\psi_-(t)}$, which embodies
both the information of the system $S$ via $S^+$ and that of the
bath spins via $I^+_{j_1}I^-_{j_2}$. We choose $j_1,~j_2$ as a pair
of nearest-neighbored sites with opposite initial spins.
Time-dependent density matrix renormalization group
method\cite{Daley-White} is employed for the calculation of
$G_{j_1,j_2}(t)$.

\begin{figure}[htbp]
\begin{center}
    \includegraphics[width=3in]{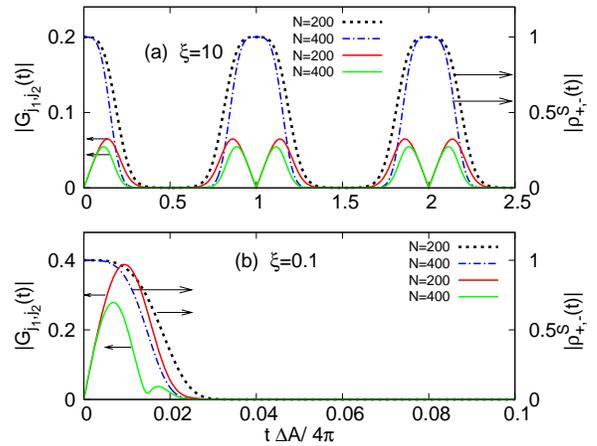}
\end{center}
 \caption[]
{ (color online). Joint-correlation evolution $|G_{j_1,j_2}(t)|$ for
(a) $\xi=10$ (b) $\xi=0.1$ with $N=200,~400$ and $j_1=N/2$,
$j_2=N/2+1$. Corresponding CE $|\rho^S_{+,-}(t)|$ are plotted for
comparison. In (a), $|G_{j_1,j_2}(t)|$ is periodic with period
$\pi$. The zero-value regimes coincide with those for CE. In (b),
after a peak close to $t=0$, $|G_{j_1,j_2}(t)|$ dies away
completely. } \label {fig:IIS}
\end{figure}

For linear HF-type interaction, we normalize the time variable $t$
by $4/\Delta A$. Our calculation shows that, the joint-correlation
evolution exhibits the same two kinds of qualitatively different
evolutions as the corresponding CE's of $S$,\cite{Wang} a periodic
evolution and a decaying evolution, see Fig.~\ref{fig:IIS}. For
$\xi=10$, $G_{j_1,j_2}(t)$ is typically periodic with period $\pi$
while for $\xi=0.1$ it decays monotonically after a narrow peak
close to $t=0$. As the chain size increases, the zero-value
intervals in the periodic evolution keep extending and the peak
width shrinking while the peak in the decaying evolution moves
closer to $t=0$.

It is interesting to notice that,
the zero-value time regimes for
$G_{j_1,j_2}(t)$ coincide with those in CE of $S$, not only in the
decaying evolution, but also in the periodic evolution. This is
excellently in consistence with Coleman-Hepp's argument for the
quantum measurement theory.\cite{Hepp} In the zero-coherence regime,
the two pointer states $\ket{\psi_\pm(t)}$ ``remain orthogonal after
any operation involving only finitely many lattice points" in the
limit of infinite chain length. We have here $I^+_{j_1}$,
$I^-_{j_2}$ as the local operators.

{\sl Pairwise concurrence of the spin chain.---}
We next investigate
the nearest-neighbored pairwise entanglement
of the environment in correspondence to the
CE of $S$.
The dynamical entanglement of the spin chain is driven by not only
the intra-bath spin-spin interaction, but also the inhomogeneous
interaction between $S$ and $E$. For $I=\frac{1}{2}$ spins, the {\it
concurrence} $\cal C$ gives a proper measure of the amount of the
pairwise entanglement,\cite{Wootters_1998} which varies from ${\cal
C}=0$ for a separable state to ${\cal C}=1$ for a maximally
entangled state.

The pairwise concurrence for chain sites $j_1$, $j_2$
can be calculated from
the corresponding reduced density matrix
$\rho_{j_1j_2} = Tr_S[Tr_{{j\!\!\backslash}\!_1 {j\!\!\backslash}\!_2 } \rho(t)]$,
where $Tr_{{j\!\!\backslash}\!_1 {j\!\!\backslash}\!_2 }$
denotes tracing over the spin chain except sites $j_1$ and $j_2$.
Noting that $I^z=\sum_{j=1}^N I^z_j$ is a good quantum number,
$\rho_{j_1j_2}$ takes the form
\begin{eqnarray}
\rho_{j_1j_2}=
\begin{pmatrix}
  u_1 & 0 & 0 & 0 \\
  0 & w_1 & z & 0 \\
  0 & z^* & w_2 & 0 \\
  0 & 0 & 0 & u_2
\end{pmatrix}
\label{eq:matrix_rdm}
\end{eqnarray}
in the standard basis
$\ket{\up_{j_1}\up_{j_2}},\ket{\up_{j_1}\down_{j_2}},
\ket{\down_{j_1}\up_{j_2}}, \ket{\down_{j_1}\down_{j_2}}$,
and the corresponding concurrence is\cite{Wootters_2001}
\begin{eqnarray}
{\cal C}(t)=2\max\Bigl[\,0, |z|-\sqrt{u_1u_2}\,\Bigr]\;.
\end{eqnarray}
The entities of matrix (\ref{eq:matrix_rdm}) are
two-point correlation functions.
In the fermionic representation their
exact expressions can be derived as
$
u_1=|c_1|^2\det[\gamma^{(+)}]+|c_2|^2\det[\gamma^{(-)}],~
u_2=1-Tr\big[|c_1|^2\gamma^{(+)}+|c_2|^2\gamma^{(-)}\big]+u_1
$
and
$
z=|c_1|^2\gamma^{(+)}_{j_1,j_2}+|c_2|^2\gamma^{(-)}_{j_1,j_2}\;,
$
where
$
\gamma^{(\pm)}_{j,j'}=\sum_{j_\beta\in\{j_\alpha\}}
\big[e^{-ih^\pm t}\big]_{j,j_\beta}
\big[e^{ih^\pm t}\big]_{j_\beta,j'}
$
with $j,~j'\in\{j_1,j_2\}$.

\begin{figure}[htbp]
\begin{center}
    \includegraphics[width=3in]{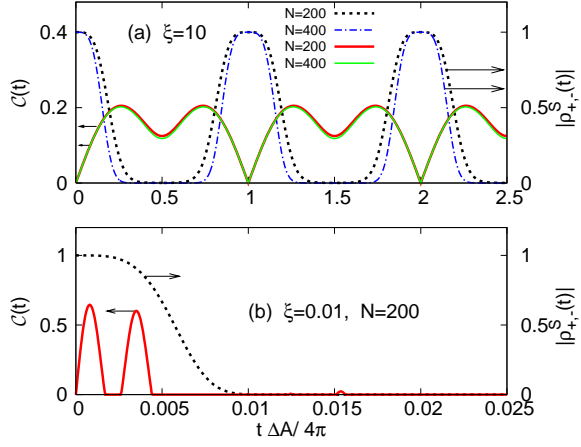}
\end{center}
 \caption[]
{ (color online). Solid lines are concurrence evolution ${\cal
C}(t)$ for (a) $\xi=10$ (b) $\xi=0.01$ with $j_1=N/2$, $j_2=N/2+1$.
The electron spin initial state is taken as $c_1=1/\sqrt 3$, and
$c_2=\sqrt 6/3$. Dotted lines are Corresponding plotted for
comparison. In (a), $\mathcal C(t)$ for $N=200$, and $N=400$ almost
overlap with each other. The evolutions are periodic with period
$\pi$ and keep non-zero value except at $t=n\pi$, $n=0,1,2\ldots\;$
In (b), $N=200$, after two narrow peaks close to $t=0$, $\mathcal
C(t)$ dies away completely. } \label {fig:Concurrence}
\end{figure}

For linear HF-type interaction, for case of $\Delta A$ much larger
than $B$, the transverse correlation $z=<I^+_{j_1}I^-_{j_2}>$
dominates the longitudinal parts $u_{1,2}=<(1/2\pm I^z_{j_1})(1/2\pm
I^z_{j_2})>$. The concurrence takes value $2|z|$ and is also
periodic with the same period as that of CE of $S$, see
Fig.~\ref{fig:Concurrence} (a). It takes zero values only at
$t=n\pi,~n=0,1,2\cdots$ and keeps non-zero within each period. For
small $\xi$, the longitudinal correlation overwhelms its transverse
counterpart, the pairwised bath concurrence exhibits an extremely
narrow peak close to $t=0$ and then dies away completely, see
Fig.~\ref{fig:Concurrence} (b).

We notice that, in the periodic CE, the zero-coherence interval in
each period extends with the increase of the chain length, see the
dashed lines in Fig.~\ref{fig:Concurrence} (a). In the limit
$N\to\infty$, it predominates the whole evolution and the periodical
coherence revival shrinks into instantaneous pulses with zero width.
This was often understood as a complete
decoherence.\cite{Sun_1993,Milburn} 
However, the bath concurrence $\mathcal C$ maintains non-zero in
these zero-coherence time intervals, and is chain-length
independent, as shown by the solid lines in
Fig.~\ref{fig:Concurrence} (a). In other words, even the periodic
coherence evolution is predominated by zero value, the bath still
keeps in a stable entangled state. Taking into consideration of the
periodically reviving coherence evolution, this persistently
entangled bath state suggests that the whole universe is in a
coherence preserving phase. 

\begin{figure}[htbp]
\begin{center}
    \includegraphics[width=3in]{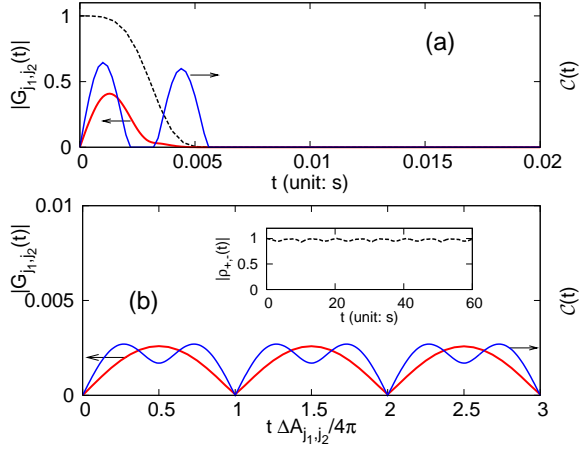}
\end{center}
\caption { (color online). Joint-correlation and bath concurrence
evolution for $A_j=\mathcal A \cos^2\frac{(j-1)\pi}{2N}$, $N=200$
with ${\cal A}=10^4~s^{-1}$, $j_1=N/2$, $j_2=N/2+1$. Dotted lines
are corresponding CE's plotted for comparison. (a) For $B=10^3$
$s^{-1}$, both $|G_{j_1,j_2}(t)|$ and $\mathcal C (t)$ exhibit zero
value after a narrow peak/peaks close to $t=0$. The corresponding
$|\rho^S_{+,-}(t)|$ decays monotonically. (b) For $B=0.1~s^{-1}$,
$|G_{j_1,j_2}(t)|$ and $\mathcal C (t)$ both exhibit periodic
evolution. The period is $\pi$ with time normalized by local
difference of the HF-type coupling strength $\Delta A_{j_1,j_2}$.
The coherence (in inset) exhibits a non-decaying irregularly
oscillating evolution. } \label {fig:Ahf_Cos}
\end{figure}

{\sl HF-type interaction in cosine form.---} Now we consider the
HF-type interaction in cosine form. The joint-correlation and bath
concurrence again exhibits two types of evolutions, see
Fig.~\ref{fig:Ahf_Cos}. When the intra-bath interaction $B$ is
large, these two quantities exhibit the same decaying behaviors as
their counterparts in the case of linear HF-type interaction,
respectively. For small $B$, unlike the non-decaying irregularly
oscillating CE, interestingly, the evolutions of both
$G_{j_1,j_2}(t)$ and $\mathcal C(t)$ are still periodic. If we
normalize the time variable $t$ by the local difference of the
HF-type interaction strength at the two sites $\Delta A_{j_1,j_2}$,
the resulted period $\pi$ is exactly the same as in the periodic
evolution with linear HF-type interaction. The scale introduced by
the two bath spins picks out a periodic component hidden in the
irregularly oscillating CE of $S$ and shows up in the
joint-correlation and bath concurrence evolutions.

\begin{figure}[htbp]
\begin{center}
    \includegraphics[width=3in]{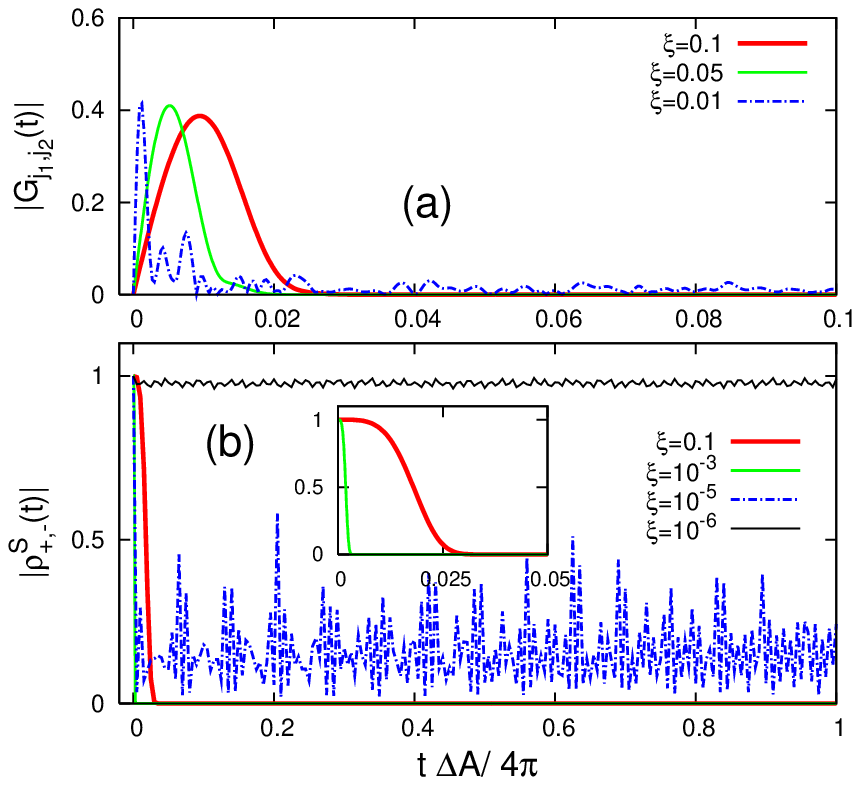}
\end{center}
\caption { (color online). $N=200$, $j_1=N/2$, $j_2=N/2+1$. (a)
Transition behavior of joint-correlation from decaying to decoupling
evolutions. For $\xi=0.1,~0.05$ lying in the parameter regime for
decaying evolutions, $|G_{j_1,j_2}(t)|$ exhibits a single peak and
then monotonically dies away. For $\xi=0.01$, the evolution starts
oscillating after the single peak, which is a kind of evidence that
the evolution is in the decoupling regime. (b) Decoupling behavior
of CE. For $\xi=0.1,~10^{-3}$ in the decoherence regime, the
coherence decays monotonically, see the inset for detail. For
$\xi=10^{-5}$, oscillation appears during the decay. The deviation
from the monotonic decay appears as a transition to the decoupling
evolution, which occurs approximately at $\xi\sim0.01/N\sim10^{-5}$.
For $\xi=10^{-6}$, as a typical decoupling behavior,
$|\rho^S_{+,-}(t)|$ fluctuates around the limit
$\lim_{\xi\to0}|\rho^S_{+,-}(t)|=1$. } \label {fig:decoupling}
\end{figure}

{\sl Decoupling phenomena---} The decoherence effect is due to the
temporally fluctuating random magnetic field exerted on the electron
spin, originated from the fluctuations of the surrounding bath spin
pairs. If the intra-bath interaction increases, the decoherence will
be enhanced owing to the increased bath spins fluctuations. However,
as can be seen from Eq.~(\ref{eq:CE}) and its relevant formula, when
the $S$-$E$ interaction becomes negligible with respect to the
intra-bath interaction, the electron spin evolves independently, in
decoupling from the bath, and its coherence takes constant value
$\lim_{\xi\to 0}\rho^S_{+,-}(t)=1$. In correspondence to the above
intuition, our calculation shows that both $\rho^S_{+,-}(t)$ and
$G_{j_1,j_2}(t)$ manifest a continuous transition from the decaying
evolution to the decoupling evolution, see
Fig.~\ref{fig:decoupling}. The bath concurrence, on the other hand,
is in association only with the quantum fluctuations of the bath
spins, although the latter is driven by the full Hamiltonian of the
universe. The decaying evolution of the bath concurrence therefore
exhibits a scaling behavior in its dependence on the intra-bath
coupling $B$ as $\mathcal C(t)\sim\mathcal C(Bt)$ while the scaling
behavior for $G_{j_1,j_2}(t)$ only appears in the decoupling regime.
The decoupling phenomenon occurs for HF-type interaction in both
linear and cosine forms, and thus is a generic feature in the large
intra-bath interaction limit.

In summary, the bath entanglement plays a significant role in
understanding the coherence status of the spin-bath system. In
particular, the subtle difference between the zero-coherence status
exhibited in two kinds of different electron spin coherence
evolutions can be discriminated by the entanglement status of the
spin chain. The periodic coherence evolution predominated by
zero-coherence and the non-decaying irregularly oscillating
coherence evolution are both associated with a periodic evolution of
the bath entanglement, which reveals the coherence-preserving nature
of the whole system and is a kind of headspring of the spin echo
effect. The zero coherence appeared in the decoherence evolution
corresponds to an environment with disentangled spin pairs. Any
disturbance to the spin on one site would have no affect to its
neighboring spins. The environment is in this sense silent and the
whole system is in a true coherence collapse state. 
As a ramification,
apparently, both the zero-coherence status
in the two different parameter regimes
meet the requirement of von Neumann's postulate .
Yet they correspond to qualitatively different
entanglement status of the environment.
This brings up a question that
whether a quantum measurement of the system
should refer to the environment status.
It is expected to stimulate further
understanding to decoherence as well as
quantum-measurement theories.


\begin{thebibliography}{12}
\bibitem{C-L}
A.O. Caldeira and A.J. Legget, Ann. Phys. (N.Y.) \textbf{149}, 374 (1983);
A.J. Leggett, S. Chakravarty, A.T. Dorsey, M.P.A. Fisher,
A. Garg, W. Zwerger, Rev. Mod. Phys. \textbf{59}, 1 (1987);
\bibitem{Hepp} K. Hepp, Hev. Phys. Acta. 45, 237 (1972); J. S. Bell, Helv. Phys. Acta, {\bf 48}, 93 (1975).
\bibitem{von-Neumann} J. von Neumann, {\sl Mathematische Grundlagen Der Quantenmechanik} (Springer, Berlin, 1932).
\bibitem{Sun_1993} C. P. Sun, \pra {\bf 48} 898 (1993).
\bibitem{Sun_2001} e.g., C. P. Sun, D. L. Zhou, S. X. Yu, X. F. Liu, Eur. Phys. J. D {\bf 13}, 145, (2001).
\bibitem{Milburn} e.g., C. M. Dawson, A. P. Hines, R. H. McKenzie, and G. J. Milburn, Phys. Rev. A \textbf{71}, 052321 (2005).
\bibitem{Wang} Zhi-Hui Wang, Bing-Shen Wang, and Zhao-Bin Su, Phys. Rev. B \textbf{78}, 054433 (2008).
\bibitem{Loss} See W. A. Coish and D. Loss, Phys. Rev. B {\bf 70}, 195340 (2004) and references therein.
\bibitem{Sham_DasSarma}   Wang Yao, Ren-Bao Liu and L. J. Sham, Phys. Rev. B \textbf{74}, 195301 (2006);
W. M. Witzel and S. Das Sarma, Phys. Rev. B \textbf{74}, 035322 (2006).
\bibitem{J-W} E. Lieb, T. Schultz, and D. Mattis, Ann. Phys. (N.Y.) {\bf 16}, 407 (1961).
\bibitem{Daley-White} Daley, A. J., C. Kollath, U. Schollwock, and G. Vidal, J. Stat. Mech. P04005(2004); S. R. White and A. E. Feiguin, Phys. Rev. Lett. \textbf{93}, 076401(2004).
\bibitem{Wootters_1998} W. K. Wootters, Phys. Rev. Lett. \textbf{80}, 2245 (1998).
\bibitem{Wootters_2001} K. M. O'Connor and W. K. Wootters, Phys. Rev. A {\bf 63}, 052302 (2001).

\end{thebibliography}
\end{document}